\documentclass[10pt,conference]{IEEEtran}
\usepackage{enumitem}
\usepackage{cite}
\usepackage{graphicx}
\usepackage{amsmath}
\usepackage{color}
\usepackage{colortbl}
\usepackage{xcolor}
\usepackage{hyperref}
\usepackage{verbatim}
\usepackage{amsthm}
\usepackage{amsfonts}
\usepackage{braket}
\usepackage{bbm}
\usepackage{bm}
\usepackage{booktabs}
\usepackage{multirow}
\usepackage{amssymb}
\usepackage{algorithm, algorithmic}
\usepackage{mathtools}
\usepackage{subcaption}
\usepackage[numbers,sort&compress,comma]{natbib}
\hypersetup{colorlinks=true, linkcolor=blue!80!black, urlcolor=blue!80!black, citecolor=blue!80!black}
\usepackage[nameinlink, capitalise]{cleveref}

\begin{document}
\title{Variational Quantum Algorithm Landscape Reconstruction by Low-Rank Tensor Completion}

% \\{\footnotesize \textsuperscript{*}Note: Sub-titles are not captured in Xplore and

\author{
\IEEEauthorblockN{Tianyi Hao$^1$, Zichang He$^2$, Ruslan Shaydulin$^2$, Marco Pistoia$^2$, Swamit Tannu$^1$}
\IEEEauthorblockA{$^1$Department of Computer Sciences, University of Wisconsin-Madison, Madison, WI 53706 USA}
\IEEEauthorblockA{$^2$Global Technology Applied Research, JPMorganChase, New York, NY 10017 USA}
% Emails: xxx
}

\maketitle
\begin{abstract}
Variational quantum algorithms (VQAs) are a broad class of algorithms with many applications in science and industry.
% the potential for quantum advantage in many industries. 
Applying a VQA to a problem involves optimizing a parameterized quantum circuit by maximizing or minimizing a cost function. A particular challenge associated with VQAs is understanding the properties of associated cost functions. Having the landscapes of VQA cost functions can greatly assist in developing and testing new variational quantum algorithms, but they are extremely expensive to compute. Reconstructing the landscape of a VQA using existing techniques
% , however, conventionally 
requires a large number of cost function evaluations, 
% sample simulations or a large amount of quantum resources, 
especially when the dimension or the resolution of the landscape is high. To address this challenge, 
% we examize the low-rankness of the interested ansatz landscape and 
we propose a low-rank tensor-completion-based approach for local landscape reconstruction. 
% Leveraging the compact representation of tensor approximation formats, we can efficiently mitigate 
By leveraging compact low-rank representations of tensors, our technique can overcome the curse of dimensionality and handle high-resolution landscapes. 
We demonstrate the power of landscapes in VQA development by showcasing practical applications
% We apply the proposed reconstruction approach in two novel applications 
of analyzing penalty terms for constrained optimization problems and examining the probability landscapes of certain basis states. 

% Thus, the problem (i.e., the ``curse”) of dimensionality arise when attempting to reconstruct the landscape of VQA cost functions where an associated quantum circuit includes a large number of parameters. 

\end{abstract}

\section{Introduction}
Quantum computers have the potential to accelerate a wide range of scientific and commercial applications
% , outperforming existing classical solutions
~\cite{2310.03011,herman2023quantum}.
% Among these, v
Variational quantum algorithms (VQAs) have attracted a lot of attention due to their broad applicability and low resource requirements, enabling their execution on near-term noisy devices~\cite{cerezo2021variational}. 
% show promise in solving a wide range of optimization problems and quantum systems, utilizing existing quantum hardware in the current near-term intermediate-scale quantum (NISQ) era~\cite{bharti2022noisy}. 
Applying a VQA to a given problem requires implementing multiple components, namely a quantum operator representing the problem of interest (\emph{Hamiltonian}), a parameterized quantum circuit (\emph{ansatz}), a classical optimizer, a cost function to minimize, an initialization strategy for the parameters, and oftentimes a strategy for taking and postprocessing measurements to mitigate noise or handle problem constraints.
In a VQA, a classical optimizer iteratively updates the circuit parameters to minimize a cost function value.
% , which is calculated with the problem Hamiltonian and measurements of the current quantum circuit. 
Then the quantum state prepared by the parameterized circuit with optimized parameters % resulting parameters found by VQAs 
correspond to high-quality solutions to the target problem.

% VQAs are expected to play a key role in the development of practical quantum computing applications. 
% However, t
The performance of a VQA depends crucially on the choice and configuration of the components described above~\cite{he2023alignment,he2023distributionally, hao2022exploiting, 10.1145/3584706, shaydulin2019multistart, Sureshbabu2023,Liu2022}. 
A poor choice of any one component can drastically decrease the algorithm's efficacy. 
Unfortunately, VQAs are difficult to analyze, tune, and debug due to their heuristic nature and complex structure.  %, making it challenging to identify the misconfigured component. 
% In addition, unique properties pertaining to quantum mechanics further limit the diagnosis of VQAs. Probing or copying the intermediate state of a quantum circuit is not possible due to the no-cloning theorem. 
% Destructive and statistical reads of the circuit output, along with noisy operations, make it demanding to obtain an accurate output. 
The need to obtain sufficient statistics on the circuit output and the presence of hardware noise contribute to the high resource requirements associated with VQA debugging. As a consequence, researchers typically try a few configurations in an ad-hoc manner to identify the 
% Traditionally, VQA researchers simply test out various configurations to derive the 
best-performing setup, which often leads to suboptimal performance.
% is highly empirical and ineffective.

The landscape of cost function values given by a range of parameter values can provide much more information than just the cost values from a single optimization~\cite{hao2023enabling,rudolph2021orqviz,perez2024analyzing}. 
In machine learning, loss function landscapes play a pivotal role in the development and fine-tuning of models~\cite{li2018visualizing}.
Similarly, VQAs can benefit from analyzing landscapes as they also have an optimizer-driven outer loop. By analyzing and visualizing landscapes in conjunction with the optimizer trace, researchers can gain insights into the optimization process. They can identify sources of misconfiguration and suboptimal behavior of VQAs and adjust the components accordingly to improve performance. Furthermore, these landscapes can reveal the presence of local minima, saddle points, and flat regions, which are critical for understanding the behavior of VQAs.

However, obtaining VQA landscapes can be extremely expensive. The naive way of generating a landscape is by performing a grid search, which involves computing the value of every point on a parameter grid. The number of points in the grid is exponential in the number of parameters of the landscape, rendering this method intractable for all but the smallest instances. 
% In addition, each cost function value requires running quantum circuits many times, which is costly to evaluate even without the large amount. In real experiments, it can take thousands of executions to calculate a single point on the landscape using near-term quantum devices.

Recently, studies have proposed using compressed sensing to reconstruct VQA landscapes using only a small number of evaluations~\cite{hao2023enabling,fontana2022efficient,lee2021progress,kim2022quantum}. This class of methods leverages the observation that the VQA landscapes are sparse in the frequency domain under Fourier transform. While such techniques have been shown to be highly accurate and stable, their applicability is limited.
% it has its limitations. 
Since compressed sensing relies on data exhibiting symmetry in the time domain, the landscape range needs to be reasonably global to exploit periodicities in VQAs. Specifically, when we want to look at details in local regions, the zoomed-in resolution is often inadequate while local regions around the optimal points are always of more interest in the practical usage of VQAs. More critically, the landscape is represented as a dense tensor of unknowns or values during or after compressed sensing~\cite{hao2023enabling}, which incurs an exponential memory and computational cost. Thus, existing techniques can only be performed on relatively low-dimensional and low-resolution landscapes, restricting the domain of applicability.

In this work, we introduce a technique for VQA landscape reconstruction from a small number of samples using low-rank tensor completion. To motivate our technique, we show extensive evidence that local VQA landscapes can be well-approximated by low-rank tensors.
% under tensor decompositions. 
% Based on this observation, we propose using tensor networks as sparse representations for VQA landscapes and utilizing tensor completion methods to reconstruct landscapes from a small number of samples. 
We implement the proposed workflow as a user-friendly, highly configurable, open-source Python package, available at \href{https://github.com/QUEST-UWMadison/OSCAR}{https://github.com/QUEST-UWMadison/OSCAR}. 
We demonstrate the power and broad applicability of our technique %We perform 
by performing numerical experiments with different VQA constructions applied to optimization and chemistry problems. % to show the effectiveness of our method. 
We identify novel applications of landscape reconstruction,
% in addition to the ones mentioned in~\cite{hao2023enabling}, 
including the analysis of penalty terms for constrained problems and insights with the probability landscapes of basis states.

\section{Background}
\subsection{Variational Quantum Algorithms}
Variational quantum algorithms (VQAs) are a class of algorithms that leverage classical optimization techniques to train parameterized quantum circuits $\ket{\Psi(\bm{\theta})}$ (\emph{ansatz}) such that the quantum state obtained by the circuit optimized parameters $\bm{\theta}$ corresponds to high-quality solutions to the target problem. The dimensionality of $\bm{\theta}$ can be generally high in order to enable the expressivity of the ansatz. As the VQAs are known to suffer from barren plateau issues and contain lots of local optimum~\cite{mcclean2018barren, 2309.07902}, a good initialization of $\bm{\theta}$ and a carefully chosen range for parameter search are necessary to enable high-quality solutions. This provides the need for
% Therefore, for the practical usage, we are usually interested in the
high-resolution cost landscapes under a local range of parameters. 

The variational quantum eigensolver (VQE) is a VQA designed for finding the ground state of a given molecule \cite{peruzzo2014variational}, and has been generalized with various ansatzes to solve a wide range of problems \cite{cerezo2021variational}. The unitary coupled-cluster singles and doubles (UCCSD) ansatz~\cite{barkoutsos2018quantum} is a chemistry-inspired VQE ansatz suitable for solving quantum chemistry problems. 

The quantum approximate optimization algorithm (QAOA)~\cite{Hogg2000,farhi2014quantum} can be viewed as a VQA with a problem-dependent ansatz. Specifically, QAOA ansatz is given by
\begin{equation}
    \ket{\Psi(\bm{\theta})}_{\text{QAOA}} = \prod_{j=1}^p\left[e^{-i\beta_j \bm{H_B}}e^{-i\gamma_j \bm{H_C}} \right]\ket{+},
\end{equation}
where $p$ is the number of layers, $\beta_1,\ldots, \beta_p$ and $\gamma_1,\ldots, \gamma_j$ are free parameters, $\bm{H_B}$ is the mixing Hamiltonian 
%(typically, a sum of single-qubit Pauli $X$ operators, $\bm{H_B}= \sum_k X_j$) 
and $\bm{H_C}$ is the Hamiltonian encoding the objective function to be optimized. QAOA has been shown to provide an asymptotic algorithmic speedup on some problems~\cite{2208.06909,shaydulin2023evidence}, motivating its study. 

\subsection{Tensor Networks}
% Tensor networks are a powerful computational tool widely used in Mathematics, Physics, Chemistry, and Computer Science~\cite{orus2014practical, vidal2003efficient, ibrahim2022constructing, he2021high, pang2020efficient, hao2022quantum}. Below, we provide a brief review of tensor networks, focusing on the concepts used in this paper.

Tensors are multi-dimensional arrays that generalize vectors and matrices. An index or bond of a tensor is one of its dimensions, and the term \emph{bond dimension} refers to the size of that dimension. For example, a vector $\bm{v} \in \mathcal{R}^{n}$ (order-1 tensor) has one index with bond dimension $n$, and a matrix $\bm{M}\in \mathcal{R}^{m\times n}$ (order-2 tensor) has two indices with bond dimensions $m$ and $n$. The multiplication between vectors and matrices can be viewed as a summation over the shared index: $\bm{M}\bm{v}=\sum_i M_{:,i}v_i$. The generalization of multiplication is tensor contraction, which sums over the shared tensor indices.

A tensor network~\cite{orus2014practical, vidal2003efficient, ibrahim2022constructing, he2021high, pang2020efficient, hao2022quantum} is a generalized graph (with dangling edges) of tensors connected by shared indices. In particular, the 1D chain tensor networks, known as Tensor Train (TT)~\cite{oseledets2011tensor} in Mathematics, or Matrix Product States (MPS)~\cite{schollwock2011density} in Physics and Chemistry, are very successful in various applications. %related to dense tensor decompositions~\cite{sidiropoulos2017tensor} and finding the ground state energy of quantum many-body systems with renormalization group methods~\cite{chan2011density}. 
A $d$-site MPS consists of 2 order-2 tensors $\mathbb{M}^{(1)},\mathbb{M}^{(d)}$ and $d-2$ order-3 tensors $\mathbb{M}^{(2)},\cdots\mathbb{M}^{(d-1)}$, such that the full contraction
% \begin{align}
%     \sum_{r_1 r_2 \ldots r_{d-1}}m^{(1)}_{i_1 r_1}m^{(2)}_{i_2 r_1 r_2}\cdots m^{(d-1)}_{i_{d-1} r_{d-2} r_{d-1}} m^{(d)}_{i_d r_{d-1}}
% \end{align}
\begin{equation}\label{eq:tt_represent}
\begin{aligned}
& \sum_{{r_1}=1}^{R_1}\sum_{{r_2}=1}^{R_2}\ldots\sum_{{r_{d-1}}=1}^{R_{d-1}} \mathbb{M}^{(1)}\left[i_1,r_1\right] \mathbb{M}^{(2)}\left[r_1,i_2,r_2\right] \cdots \\
&  \mathbb{M}^{(d-1)}\left[r_{d-2},i_{d-1},r_{d-1}\right] \mathbb{M}^{(d)}\left[r_{d-1},i_d\right]
\end{aligned}
\end{equation}
gives the dense tensor it represents. 
% The ``open" indices $i_1,\cdots,i_d$ that are uncontracted are called \emph{physical indices}, which usually correspond to basis states of a quantum system, discrete variables, or parameters. The interconnected indices $r_1,r_2,\cdots,r_{d-1}$ are called \emph{virtual indices} or \emph{bond indices} and their maximum value $R_1,R_2,\cdots,R_{d-1}$ are called TT ranks or bond dimensions.
The bond dimensions $R_1,R_2,\cdots,R_{d-1}$ of the interconnected indices $r_1,r_2,\cdots,r_{d-1}$ are referred to as \emph{ranks}.

% One of MPS's advantages over other tensor networks is its canonical form, where each tensor except for the canonical center is isometric. An isometric tensor contracts to an identity matrix with its conjugate transpose. Thus, an inner product between a canonicalized MPS and its complex conjugate is reduced to the contraction of the canonical center. Canonicalization can be done by sequential local contraction and SVD or QR decomposition pairs.

% Tensor diagrams are an intuitive way of visualizing tensors and tensor networks. Each node represents a tensor, and each edge represents an index. By convention, a triangular shape denotes the tensor is isometric. Figure TOADD shows the tensor diagram of an MPS, its left canonical form, and the reduction of an inner product.
\begin{figure}[t]
    \centering
    \includegraphics[width = \linewidth]{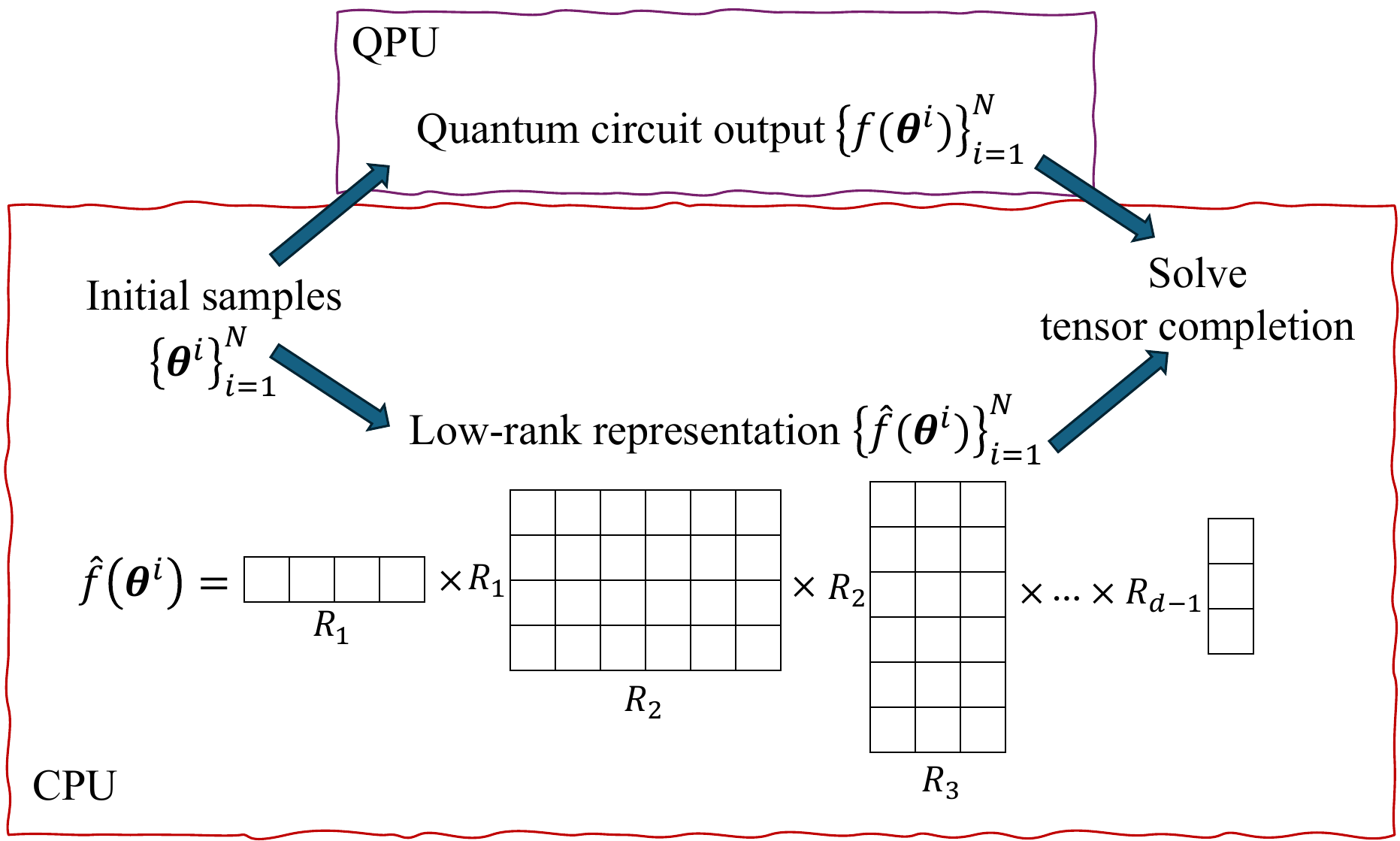}
    \caption{Overview of the proposed landscape reconstruction method by low-rank tensor completion. A small number of quantum circuit outputs are used to perform tensor completion on the low-rank representation of the landscape.
    }
    \label{fig:overview}
\end{figure}

\section{Landscape Reconstruction by Low-rank Tensor Completion}
Without loss of generality, the cost function of a VQA $f(\bm{\theta})$ can be defined as $\braket{\Psi(\bm{\theta})|\bm{O}|\Psi(\bm{\theta})}$, where $\bm{O}$ is some observable of interest. Generally, $\bm{\theta}$ is a $d$-dimensional continuous variable. To enable landscape reconstruction, we first discretize $\bm{\theta}$ using a $d$-dimensional grid, which discretizes the $i$-th parameter into $N_i$ points.
% For the ease of landscape reconstruction,  $\bm{\theta}$ is discretized into finite resolutions $\{N_i\}_{i=1}^d$. 
As a consequence of the curse of dimensionality, the space (memory) cost of landscape grows exponentially with $d$ as $\prod_{i=1}^d N_i$
% $\mathcal{O}(N^d)$ 
% Meanwhile, this cost can quickly become unaffordable when we handle a high-resolution landscape with large $\{N_i\}_{i=1}^d$.
Storing the grid explicitly is impossible for even moderate $d$.

Given a $d$-dimensional VQA parameter $\bm{\theta} = \left(\theta_1, \ldots, \theta_d \right)$ with indices $\left(i_1, \ldots, i_d \right)$, $f(\bm{\theta})$ can be approximated by Eq.~\eqref{eq:tt_represent}. {\em The space complexity of a TT representation is $R_1 N_1 + R_{d-1} N_d + \sum_{i=2}^{d-1} R_{i-1} R_i N_i$, reducing the memory requirements from exponential to linear with respect to $d$ and $N_i$}. The compact TT representation accurately expresses the landscape as long as it is low rank. We show evidence that the landscape is low-rank in~\cref{sec:low-rank}.
% \begin{equation}
% \begin{aligned}
% & \hat{f}(\theta_1, \theta_2, \ldots, \theta_d) =  
%     \sum_{{r_1}=1}^{R_1}\sum_{{r_2}=1}^{R_2}\ldots\sum_{{r_{d-1}}=1}^{R_{d-1}} \\
%   &  m^{(1)}_{{i_1}{r_1}} m^{(2)}_{{i_2}{r_1}{r_2}} \ldots  m^{(d-1)}_{{i_{d-1}}{r_{d-2}}{r_{d-1}}} m^{(d)}_{{i_d}{r_{d-1}}}
% \end{aligned}
% \end{equation}
% \begin{equation}
% \begin{aligned}
% & \hat{f}(\theta_1, \theta_2, \ldots, \theta_d) =  
%     \sum_{{r_1}=1}^{R_1}\sum_{{r_2}=1}^{R_2}\ldots\sum_{{r_{d-1}}=1}^{R_{d-1}} \\
% & \mathbb{M}^{(1)}\left[i_1,r_1\right] \mathbb{M}^{(2)}\left[r_1,i_2,r_2\right] \cdots \\
% &  \mathbb{M}^{(d-1)}\left[r_{d-2},i_{d-1},r_{d-1}\right] \mathbb{M}^{(d)}\left[r_{d-1},i_d\right]
% \end{aligned}
% \end{equation}
Here, we use the TT format, but other tensor network and tensor decomposition formats are generally applicable. 

The proposed landscape reconstruction process is illustrated in \cref{fig:overview}. To begin with, we need $N$ samples from the $(N_1, N_2 , \ldots, N_d)$-sized full landscape tensor ($N \ll \prod_{i=1}^d N_i$) and execute the associated circuits in a quantum computer to obtain $\bm{f}(\bm{\theta}) = \{f (\bm{\theta}^i)\}_{i=1}^N$. Next, we create the data structure for low-rank representation $\bm{\hat{f}}(\bm{\theta})=\{\hat{f} (\bm{\theta}^i)\}_{i=1}^N$ and solve a tensor completion problem through optimizing low-rank factors $\{\mathbb{M}^{(i)}\}_{i=1}^d$. After the tensor completion, given any unsampled parameter of the ansatz, we can predict its output through the full contraction given by~\cref{eq:tt_represent}.

\subsection{Tensor Completion Formulation}
The tensor completion problem for the tensor-train format tensor is formulated as~\cite{grasedyck2015alternating}:
\begin{equation}~\label{eq:completion_problem}
   \min_{\{\mathbb{M}^{(i)}\}_{i=1}^d} \|\bm{f}(\bm{\theta}) - \hat{\bm{f}}(\bm{\theta})\|_F
\end{equation}
Here, we mainly use the algorithm in Ref.~\cite{chertkov2023black} to solve the low-rank tensor completion problem, where the initial approximation of the factor matrices $\{\mathbb{M}^{(i)}\}_{i=1}^d$ are set based on an ANOVA (analysis of variance) approach and then optimized in an ALS (alternating least square) algorithm. We note that alternative algorithms~\cite{yuan2019high,steinlechner2016riemannian} for solving tensor completion problems can also be used in our landscape reconstruction procedure. 

\subsection{Evidence that Local VQA Landscapes are Low-rank}\label{sec:low-rank}
To verify that the local VQA landscapes are low-rank, we obtain the landscapes of various configurations in the dense tensor format by numerical simulation. With the dense landscapes, we perform sequential reshaping and singular value decompositions (SVD) along each dimension of the dense tensor to transform it into the TT format. This process is known as the TT decomposition, where the approximation error can be upper bounded by the truncation threshold of the singular values during SVDs. The number of remaining singular values after truncation is the aforementioned \emph{rank} of the TT. 
% A tensor train of $k$ tensors with physical dimension $d$ and ranks $R_1,R_2,\ldots,R_{d-1}$ takes space of size $\sum_{i=1}^{k} dr_{i-1}r_{i}$, where $r_0 = r_k = 1$, which is polynomial if the ranks are controllable. In comparison, the corresponding $k$-dimensional dense tensor with size $d$ for each dimension takes space of exponential size $d^k$.

\begin{table*}[ht]
\centering
% \resizebox{\columnwidth}{!}{
\begin{tabular}{cc cc cc  cccc cccc  ccc ccc}
\toprule
\multicolumn{2}{c}{VQA Ansatz} & \multicolumn{4}{c}{QAOA $p=1$} & \multicolumn{8}{c}{QAOA $p=2$} & \multicolumn{6}{c}{UCCSD H$_2$}\\
\cmidrule(lr){1-2}\cmidrule(lr){3-6}\cmidrule(lr){7-14}\cmidrule(lr){15-20}
\multicolumn{2}{c}{Truncation threshold} & \multicolumn{2}{c}{$10^{-2}$} & \multicolumn{2}{c}{$10^{-5}$} & \multicolumn{4}{c}{$10^{-2}$} & \multicolumn{4}{c}{$10^{-5}$} & \multicolumn{3}{c}{$10^{-2}$} & \multicolumn{3}{c}{$10^{-5}$}\\
\cmidrule(lr){1-2}\cmidrule(lr){3-4}\cmidrule(lr){5-6}\cmidrule(lr){7-10}\cmidrule(lr){11-14}\cmidrule(lr){15-17}\cmidrule(lr){18-20}
Range & Resolution & Rank & Space & Rank & Space & \multicolumn{3}{c}{Rank} & Space & \multicolumn{3}{c}{Rank} & Space & \multicolumn{2}{c}{Rank} & Space & \multicolumn{2}{c}{Rank} & Space \\
\cmidrule(lr){1-2}\cmidrule(lr){3-4}\cmidrule(lr){5-6}\cmidrule(lr){7-10}\cmidrule(lr){11-14}\cmidrule(lr){15-17}\cmidrule(lr){18-20}
\multirow{3}{*}{$\frac{\pi}{4}$} 
& 16   & 2 & 4$\times$ & 2 & $4\times$   & 4 & 6 & 3 & $84\times$  & 7 & 13 & 6 & 23$\times$   & 2 & 2 & 32$\times$  & 3 & 3 & 17$\times$ \\
& 32   & 2 & 8$\times$ & 2 & $8\times$   & 4 & 5 & 3 & $780\times$  & 7 & 12 & 6 & 194$\times$   & 2 & 2 & 128$\times$  & 3 & 3 & 68$\times$ \\
& 64   & 2 & 16$\times$ & 2 & $16\times$   & 4 & 6 & 3 & $5350\times$  & 7 & 13 & 6 & 1440$\times$   & 2 & 2 & 512$\times$  & 3 & 3 & 273$\times$ \\
\cmidrule(lr){1-2}\cmidrule(lr){3-4}\cmidrule(lr){5-6}\cmidrule(lr){7-10}\cmidrule(lr){11-14}\cmidrule(lr){15-17}\cmidrule(lr){18-20}
\multirow{3}{*}{$\frac{\pi}{8}$} 
& 16   & 1 & 8$\times$ & 2 & $4\times$   & 2 & 3 & 3 & $205\times$  & 5 & 9 & 4 & 46$\times$   & 1 & 1 & 85$\times$  & 3 & 3 & 17$\times$ \\
& 32   & 1 & 16$\times$ & 2 & $8\times$   & 2 & 3 & 2 & $2048\times$  & 5 & 9 & 4 & 364$\times$   & 1 & 1 & 341$\times$  & 3 & 3 & 68$\times$ \\
& 64   & 1 & 32$\times$ & 2 & $16\times$   & 2 & 3 & 2 & $16384\times$  & 5 & 9 & 4 & $2913\times$   & 1 & 1 & 1365$\times$  & 3 & 3 & 273$\times$ \\
\cmidrule(lr){1-2}\cmidrule(lr){3-4}\cmidrule(lr){5-6}\cmidrule(lr){7-10}\cmidrule(lr){11-14}\cmidrule(lr){15-17}\cmidrule(lr){18-20}
\multirow{3}{*}{$\frac{\pi}{16}$} 
& 16   & 1 & 8$\times$ & 2 & $4\times$   & 2 & 2 & 2 & 341$\times$  & 4 & 6 & 4 & 73$\times$   & 1 & 1 & 85$\times$  & 3 & 2 & 23$\times$ \\
& 32   & 1 & 16$\times$ & 2 & $8\times$   & 2 & 2 & 1 & 3641$\times$  & 4 & 6 & 4 & 585$\times$   & 1 & 1 & 341$\times$  & 3 & 2 & 93$\times$ \\
& 64   & 1 & 32$\times$ & 2 & $16\times$   & 2 & 2 & 1 & $29127\times$  & 4 & 6 & 4 & $4681\times$   & 1 & 1 & 1365$\times$  & 3 & 2 & 372$\times$ \\
\bottomrule
\end{tabular}
% }
\caption{Rank and space reduction of representing VQA landscapes in the TT format with specified truncation thresholds. Each row shows a different landscape range and resolution combination.}
\label{tab:rank}
\end{table*}

\cref{tab:rank} summarizes the rank and space reduction results by performing the TT decomposition of a fully sampled landscape tensor. We performed one-and-two-layer QAOA solving the Maximum Cut (MaxCut) problem and VQE with the UCCSD ansatz solving the Hydrogen molecule. 

For each VQA configuration, we show the ranks and space reduction after truncating singular values that are less than $10^{-2}$ and $10^{-5}$ of the 2-norm of all singular values, respectively. 
For simplicity, we let each dimension share the same resolution. The interested parameter space is defined as $\left[ \theta_{\text{center}} - \frac{l}{2}, \theta_{\text{center}} + \frac{l}{2} \right]$, where $\theta_{\text{center}}$ is the center of parameter space and $l$ is the range of a parameter. We test landscape ranges $l\in \{\frac{\pi}{4}, \frac{\pi}{8} ,\frac{\pi}{16}\}$ and resolutions $16$, $32$, and $64$ along each dimension. The landscape is centered at informed initial points of respective VQAs. For QAOA with MaxCut, we set $\bm{\theta}_{\text{center}}$ based on QAOA parameter concentration in~\cite{wurtz2021fixed}. For the Hydrogen molecule, we use the state given by the Hartree-Fock method. These initial points have been shown to be reasonably close to the optimal points. From practical perspectives, researchers often use them as initializations in actual VQA experiments, and local optimizations in a small surrounding region are typically enough to find the optimal point.

We observe that generally, the rank remains low compared to the resolution of the landscape, resulting in substantial space reduction. \emph{Notably, it is independent of the landscape resolution, enabling the sparse representation of very dense landscapes.} As expected, higher-dimensional landscapes have higher ranks, and more local landscapes have lower ranks.

\subsection{Evaluation}

We implement the TT representation of landscapes and the tensor completion reconstruction workflow as part of the VQA helper package OSCAR (\href{https://github.com/QUEST-UWMadison/OSCAR}{https://github.com/QUEST-UWMadison/OSCAR}). We employ the teneva package~\cite{chertkov2022optimization, chertkov2023black} for the tensor completion functionalities and provide a unified interface with the compressed-sensing-based reconstruction method in~\cite{hao2023enabling}. Note that due to the discrepancy in the applicable scope of the two methods, it is not possible to directly compare their performance.

\begin{figure*}
    \centering
    \begin{subfigure}{0.325\textwidth}
        \centering
        \includegraphics{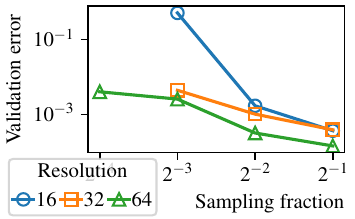}
        \subcaption{QAOA $p=1$}
    \end{subfigure}
    \begin{subfigure}{0.325\textwidth}
        \centering
        \includegraphics[trim={0.6cm, 0, 0, 0}, clip]{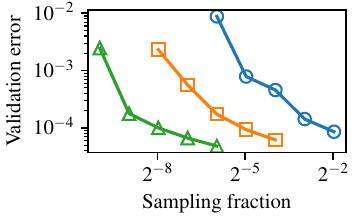}
        \subcaption{QAOA $p=2$}
    \end{subfigure}
    \begin{subfigure}{0.325\textwidth}
        \centering
        \includegraphics[trim={0.6cm, 0, 0, 0}, clip]{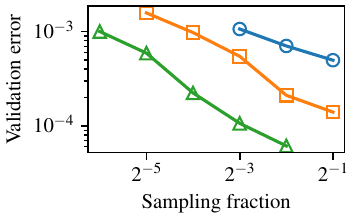}
        \subcaption{UCCSD H$_2$}
    \end{subfigure}
    \caption{Validation error as a function of sampling fraction for resolutions 16, 32, and 64. The reconstruction error drops steadily as the sampling fraction increases. Note that at the same error level, the required sampling fraction declines exponentially as the resolution doubles.}
    \label{fig:error}
\end{figure*}

We perform numerical simulation of QAOA of various depths, solving the MaxCut problem and VQE with the UCCSD ansatz solving the Hydrogen molecule. We fix the landscape range to be $\frac{\pi}{16}$ and vary the resolution and sampling fraction. We use second-order ANOVA decomposition, followed by $1000$ iterations of ALS with regularization of $0.01$. We set the rank for both ANOVA and ALS to be $2$, $6$, and $3$ for $p=1$, $p=2$ QAOA, and H$_2$, respectively. We take $5\%$ of the samples as the validation set and calculate the relative error between the true values and the reconstructed values.

\cref{fig:error} shows the validation error as a function of sampling fraction for resolutions $16$, $32$, and $64$. We see that the error drops steadily as the sampling fraction increases. Since the three problem configurations have two, four, and three dimensions respectively, doubling the resolution increases the total number of points on the landscape by $4$, $16$, and $8$ times. Importantly, note that the sampling fraction for achieving the same error also drops exponentially as the resolution doubles. Thus, the tensor-completion-based reconstruction allows us to greatly enhance the resolution of the discretization of the landscape with a remarkably small cost in comparison. Approximately, doubling the resolution only requires twice the amount of samples, instead of $2^d$ times, to keep the same level of accuracy. 

\begin{figure}[t]
    \centering
    \includegraphics[trim={1.8cm, 0, 1.8cm, 0}, width=\columnwidth, clip]{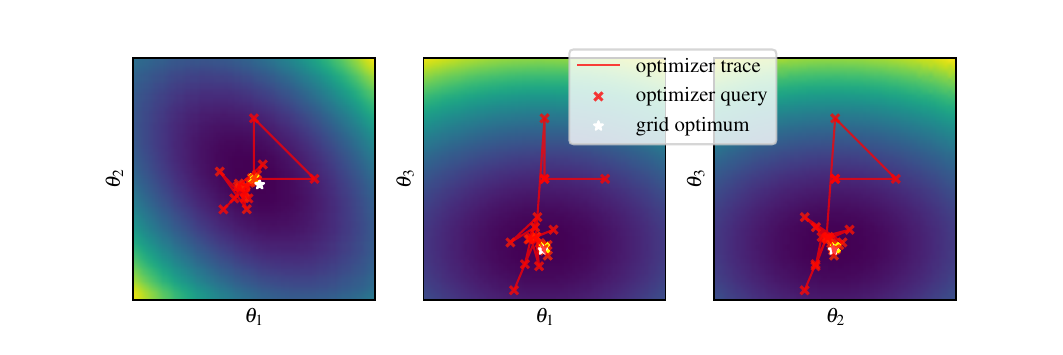}
    \caption{2D slices of the 3D landscape of UCCSD ansatz solving the Hydrogen molecule. Each figure has one of the three parameters fixed at the value given by the Hartree-Fock method. The optimization trace by the COBYLA method is projected and overlaid on each landscape slice, showing that only $\theta_3$ needs to be predominantly optimized.}
    \label{fig:h2}
\end{figure}

For a reference of the error level, \cref{fig:h2} shows an example landscape reconstruction of VQE with the UCCSD ansatz solving the Hydrogen molecule. Since the landscape is three-dimensional, we present three 2D slices where one of the parameters is fixed at the center. The landscape range is $\frac{\pi}{8}$ and the resolution is 64. The validation error is $5.4\times 10^{-4}$. Visually, there is little discrepancy between the reconstructed landscape and the exactly evaluated landscape.

With the optimizer trace overlaid, \cref{fig:h2} also demonstrates a simple use case of the landscape. We use COBYLA as the optimizer, which has one initial query in each direction, forming the triangles we see in the projected trace. The center of the landscape is the initial point of the optimization, which is derived by the Hartree-Fock method. We can clearly see that only $\theta_3$ of the three parameters needs to be optimized. For $\theta_1$ and $\theta_2$, COBYLA decides that the Hartree-Fock values are close enough to the optimal values and do not need much adjustment. Thus, in actual experiments when the number of quantum processor queries is very limited, we can fix $\theta_1$ and $\theta_2$ and only optimize for $\theta_3$. Doing so is especially helpful for optimizers assuming a complex model and requiring initial queries that are superlinear in the number of parameters.

\section{Landscape Applications}

In this section, we showcase novel examples of how landscape information can be used in VQA research and development. For additional applications % use cases that have already been documented
such as efficient benchmarking optimizers, configuring noise mitigation methods, and informed initialization, the reader is referred to Ref.~\cite{hao2023enabling}.

\subsection{Understanding Penalty Terms}

One of the central challenges in applying VQAs to many real-world optimization problems is to take into consideration their constraints. VQAs can address these constraints by enforcing them in the ansatz, but this approach often leads to deep circuits that behave poorly on near-term noisy hardware.
The simplest and most widely used approach is to introduce a penalty term $\bm{H}_P$ in the Hamiltonian, responsible for enforcing the constraint. VQAs then work with the penalized Hamiltonian $\bm{H}=\bm{H_C}+\lambda \bm{H}_P$ to balance between solution quality and solution feasibility, where $\bm{H}_C$ is the cost Hamiltonian and $\lambda$ is a penalty factor that controls the effectiveness of the penalty. VQAs rely on a carefully selected penalty factor to ensure high-quality in-constraint solutions can be found with high probability~\cite{herman2023constrained,niroula2022constrained,hao2022exploiting}. Unfortunately, like configuring other VQA components, tuning the effect of the penalty term is typically carried out in an empirical fashion, presenting a nontrivial challenge.

We can employ VQA landscapes to help analyze, visualize, and configure penalty terms. Each point of the landscape is the expectation value of the circuit and the Hamiltonian $\braket{\Psi(\bm{\theta})|\bm{H}|\Psi(\bm{\theta})}$. By the linearity of expectation, it can be decomposed to linear combinations of expectations of subterms in the Hamiltonian:
\begin{align*}
    &\braket{\Psi(\bm{\theta})|\bm{H_C}+\lambda \bm{H}_P|\Psi(\bm{\theta})} \\= &\braket{\Psi(\bm{\theta})|\bm{H_C}|\Psi(\bm{\theta})} + \lambda\braket{\Psi(\bm{\theta})|\bm{H_P}|\Psi(\bm{\theta})}.
\end{align*}

Therefore, with the cost landscape and the penalty landscape, we can trivially get arbitrary linear combinations of them. The effect of the penalty term can thus be understood by comparing landscapes with varying penalty factors. 

\begin{figure}[t]
    \centering
    \begin{subfigure}[t]{0.325\columnwidth}
    \centering
    \includegraphics[width=\columnwidth, trim={2.33cm, 0.5cm, 0.73cm, 0}, clip]{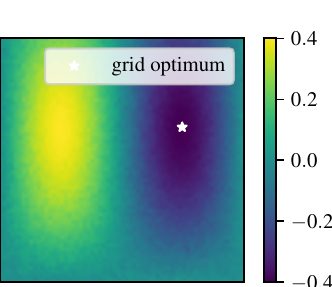}
    \subcaption{\quad\quad}
    \end{subfigure}
    \begin{subfigure}[t]{0.325\columnwidth}
    \centering
    \includegraphics[width=\columnwidth, trim={2.33cm, 0.5cm, 0.73cm, 0}, clip]{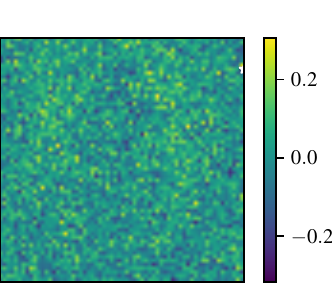}
    \subcaption{\quad\quad}
    \end{subfigure}
    \begin{subfigure}[t]{0.325\columnwidth}
    \centering
    \includegraphics[width=\columnwidth, trim={2.33cm, 0.5cm, 0.73cm, 0}, clip]{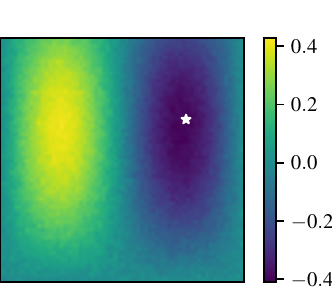}
    \subcaption{\quad\quad}
    \end{subfigure}
    \begin{subfigure}[t]{0.325\columnwidth}
    \centering
    \includegraphics[width=\columnwidth, trim={2.33cm, 0.5cm, 0.73cm, 0}, clip]{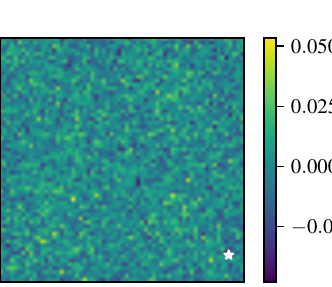}
    \subcaption{\quad\quad}
    \end{subfigure}
    \begin{subfigure}[t]{0.325\columnwidth}
    \centering
    \includegraphics[width=\columnwidth, trim={2.33cm, 0.5cm, 0.73cm, 0}, clip]{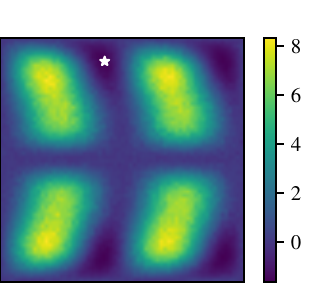}
    \subcaption{\quad\quad}
    \end{subfigure}
    \begin{subfigure}[t]{0.325\columnwidth}
    \centering
    \includegraphics[width=\columnwidth, trim={2.33cm, 0.5cm, 0.73cm, 0}, clip]{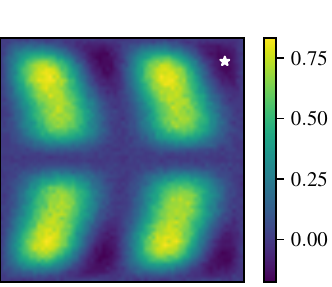}
    \subcaption{\quad\quad}
    \end{subfigure}
    \begin{subfigure}[t]{0.325\columnwidth}
    \centering
    \includegraphics[width=\columnwidth, trim={2.33cm, 0.5cm, 0.73cm, 0}, clip]{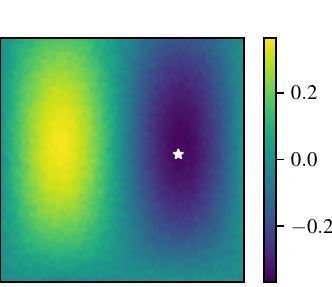}
    \subcaption{\quad\quad}
    \end{subfigure}
    \begin{subfigure}[t]{0.325\columnwidth}
    \centering
    \includegraphics[width=\columnwidth, trim={2.33cm, 0.5cm, 0.73cm, 0}, clip]{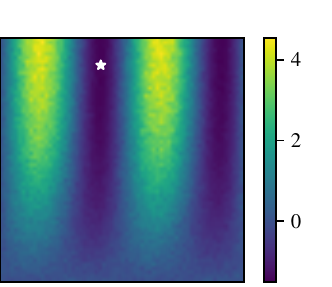}
    \subcaption{\quad\quad}
    \end{subfigure}
    \begin{subfigure}[t]{0.325\columnwidth}
    \centering
    \includegraphics[width=\columnwidth, trim={2.33cm, 0.5cm, 0.73cm, 0}, clip]{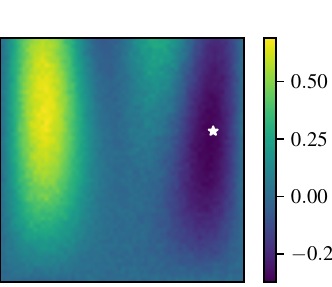}
    \subcaption{\quad\quad}
    \end{subfigure}
    \caption{
    Different combinations of the cost, penalty, and penalized Hamiltonians used in ansatz construction (rows) and as observable in the expectation (columns). That is, each row shares the same ansatz, constructed with the cost, penalty, and penalized Hamiltonians, from top to bottom. From left to right, each column uses the cost, penalty, and penalized Hamiltonians as the observable. So as an example, \cref{fig:penalty} (g) shows the landscape of the penalized ansatz with the cost Hamiltonian $\braket{\Psi_{\bm{H}}(\bm{\theta})|\bm{H}_C|\Psi_{\bm{H}}(\bm{\theta})}$.
    % (a) Ansatz built with the cost Hamiltonian and the cost Hamiltonian as the observable $\braket{\Psi_{\bm{H}_C}(\bm{\theta})|\bm{H}_C|\Psi_{\bm{H}_C}(\bm{\theta})}$; (b) ansatz built with the cost Hamiltonian and the penalty Hamiltonian as the observable $\braket{\Psi_{\bm{H}_C}(\bm{\theta})|\bm{H}_P|\Psi_{\bm{H}_C}(\bm{\theta})}$; (c) ansatz built with the cost Hamiltonian and the penalized Hamiltonian as the observable $\braket{\Psi_{\bm{H}_C}(\bm{\theta})|\bm{H}|\Psi_{\bm{H}_C}(\bm{\theta})}$; (d) ansatz built with the cost Hamiltonian and the cost Hamiltonian as the observable $\braket{\Psi_{\bm{H}_P}(\bm{\theta})|\bm{H}_C|\Psi_{\bm{H}_P}(\bm{\theta})}$; (e) ansatz built with the penalty Hamiltonian and the penalty Hamiltonian as the observable $\braket{\Psi_{\bm{H}_P}(\bm{\theta})|\bm{H}_P|\Psi_{\bm{H}_P}(\bm{\theta})}$; (f) ansatz built with the penalty Hamiltonian and the penalized Hamiltonian as the observable $\braket{\Psi_{\bm{H}_P}(\bm{\theta})|\bm{H}_C|\Psi_{\bm{H}}(\bm{\theta})}$; (g) ansatz built with the penalized Hamiltonian and the cost Hamiltonian as the observable $\braket{\Psi_{\bm{H}}(\bm{\theta})|\bm{H}_C|\Psi_{\bm{H}}(\bm{\theta})}$; (h) ansatz built with the penalized Hamiltonian and the penalty Hamiltonian as the observable $\braket{\Psi_{\bm{H}}(\bm{\theta})|\bm{H}_P|\Psi_{\bm{H}}(\bm{\theta})}$; (i) ansatz built with the penalized Hamiltonian and the penalized Hamiltonian as the observable $\braket{\Psi_{\bm{H}}(\bm{\theta})|\bm{H}|\Psi_{\bm{H}}(\bm{\theta})}$.
    }
    \label{fig:penalty}
    \vspace{-0.1in}
\end{figure}

The above method applies to cases where the ansatz does not change with the Hamiltonian. Here we show a more complicated example. For the original QAOA, whose ansatz is constructed with the knowledge of the Hamiltonian, we need to take additional considerations. \cref{fig:penalty} shows different combinations of the cost, penalty, and penalized Hamiltonians of the Portfolio Optimization problem used in ansatz construction and as observable in the expectation. From top to bottom, each row corresponds to the ansatz constructed with the cost, penalty, and penalized Hamiltonians, respectively. Similarly, the columns correspond to the Hamiltonians used as observable in the expectation. So as an example, \cref{fig:penalty} (g) shows the landscape of the penalized ansatz with the cost Hamiltonian $\braket{\Psi_{\bm{H}}(\bm{\theta})|\bm{H}_C|\Psi_{\bm{H}}(\bm{\theta})}$. The range of landscapes is chosen to cover exactly one cycle (a unique region) in the periodic parameter space.

% Based on the cost landscape (\cref{fig:penalty} (a)), we see that the optimal point, which produces the highest quality solutions (without considering the constraints), is located at the center of the right concave. However, the penalty landscape (\cref{fig:penalty} (e)) shows that most parts of the right concave yield high constraint violations and are unsuitable as produced output.
We observe that cost ansatz with penalty Hamiltonian (\cref{fig:penalty} (b)) and penalty ansatz with cost Hamiltonian (\cref{fig:penalty} (d)) are random noise-like signals, making the cost and penalty ansatzes with the penalized Hamiltonian (\cref{fig:penalty} (c) and (f)) just a noisy version of the cost landscape and the penalty landscape. This is due to the relative value of the penalty or the cost being too small, making the landscape impossible to resolve. Similar observations have motivated the rescaling of QAOA cost Hamiltonian in Refs.~\cite{10.1145/3584706,Sureshbabu2023}. Thus, for constrained problems, the cost Hamiltonian or the penalty Hamiltonian cannot be used alone when constructing the QAOA ansatz, even if the expectation observable contains both terms. 

\begin{figure}[t]
    \centering
    \begin{subfigure}[t]{0.325\columnwidth}
    \centering
    \includegraphics[width=\columnwidth, trim={2.33cm, 0.5cm, 0.73cm, 0}, clip]{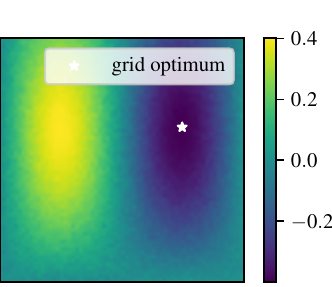}
    % \subcaption{\quad\quad}
    \end{subfigure}
    \begin{subfigure}[t]{0.325\columnwidth}
    \centering
    \includegraphics[width=\columnwidth, trim={2.33cm, 0.5cm, 0.73cm, 0}, clip]{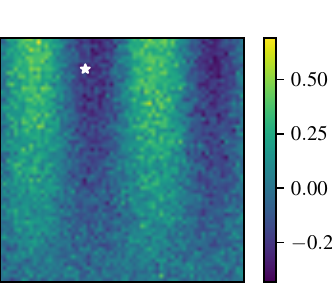}
    % \subcaption{\quad\quad}
    \end{subfigure}
    \begin{subfigure}[t]{0.325\columnwidth}
    \centering
    \includegraphics[width=\columnwidth, trim={2.33cm, 0.5cm, 0.73cm, 0}, clip]{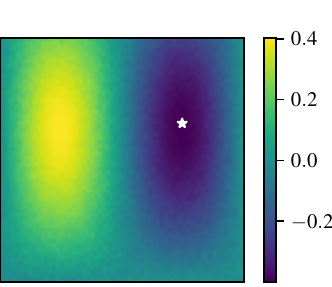}
    % \subcaption{\quad\quad}
    \end{subfigure}
    \begin{subfigure}[t]{0.325\columnwidth}
    \centering
    \includegraphics[width=\columnwidth, trim={2.33cm, 0.5cm, 0.73cm, 0}, clip]{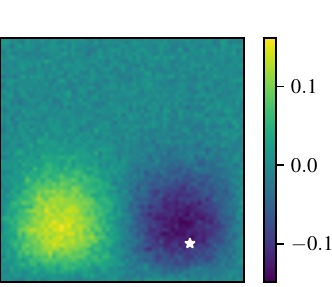}
    % \subcaption{\quad\quad}
    \end{subfigure}
    \begin{subfigure}[t]{0.325\columnwidth}
    \centering
    \includegraphics[width=\columnwidth, trim={2.33cm, 0.5cm, 0.73cm, 0}, clip]{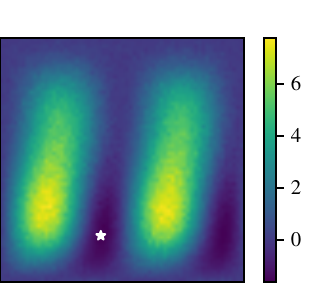}
    % \subcaption{\quad\quad}
    \end{subfigure}
    \begin{subfigure}[t]{0.325\columnwidth}
    \centering
    \includegraphics[width=\columnwidth, trim={2.33cm, 0.5cm, 0.73cm, 0}, clip]{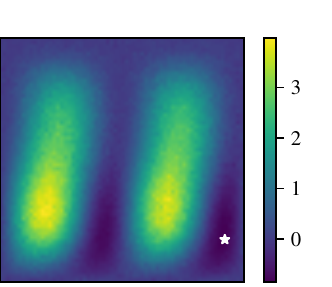}
    % \subcaption{\quad\quad}
    \end{subfigure}
    \begin{subfigure}[t]{0.325\columnwidth}
    \centering
    \includegraphics[width=\columnwidth, trim={2.33cm, 0.5cm, 0.73cm, 0}, clip]{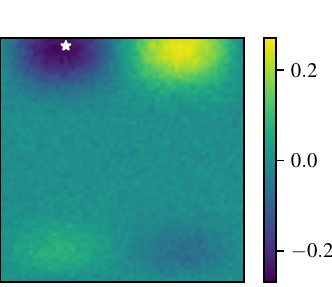}
    % \subcaption{\quad\quad}
    \end{subfigure}
    \begin{subfigure}[t]{0.325\columnwidth}
    \centering
    \includegraphics[width=\columnwidth, trim={2.33cm, 0.5cm, 0.73cm, 0}, clip]{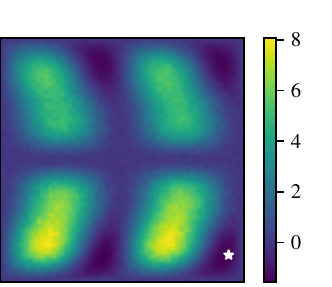}
    % \subcaption{\quad\quad}
    \end{subfigure}
    \begin{subfigure}[t]{0.325\columnwidth}
    \centering
    \includegraphics[width=\columnwidth, trim={2.33cm, 0.5cm, 0.73cm, 0}, clip]{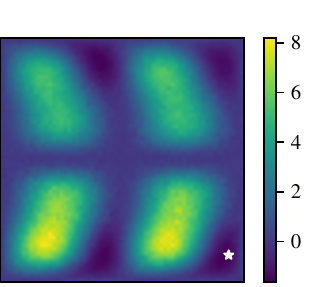}
    % \subcaption{\quad\quad}
    \end{subfigure}
    \caption{Examples showing the effect of poorly chosen penalty factors. Each row uses the same penalty factor. From the first to the third row, the penalty factors are 0.01, 0.5, and 1. From left to right, the columns show cost, penalty, and penalized Hamiltonians as the observable in expectation values.}
    \label{fig:penalty-factor}
\end{figure}

But how does an ansatz constructed with a penalized Hamiltonian achieve a balance between the two terms? Based on the magnitude of the cost and penalty landscapes, we select a penalty factor of 0.1 to bring them to the same scale. We see that the cost landscape (\cref{fig:penalty} (g)) remains approximately the same. In particular, the lowest value achievable, which corresponds to the highest solution quality possible, is not sacrificed too much. On the other hand, the penalty landscape (\cref{fig:penalty} (h)) becomes completely different, where the feasible region aligns considerably better with the cost landscape. As a result, the optimal point of the combined landscape (\cref{fig:penalty} (i)) is reasonably close to the optimal point of the cost landscape to produce high-quality solutions while being highly feasible according to the penalty landscape.

\cref{fig:penalty-factor} shows three examples where the penalty factor is poorly chosen. From the first to the third row, the penalty factors are 0.01, 0.5, and 1. From left to right, the columns are cost, penalty, and penalized landscapes. We notice that decreasing the penalty factor too much transforms the penalty landscape into a random noise-like signal while increasing the penalty factor over the balance noticeably deforms the cost landscape. The results of both are not desirable.

\subsection{Probability Landscape of Basis State}
\begin{figure}[t]
    \centering
    \begin{subfigure}[t]{0.325\columnwidth}
    \centering
    \includegraphics[width=\columnwidth, trim={2.33cm, 0.5cm, 0.73cm, 0}, clip]{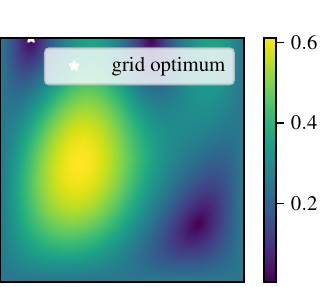}
    \subcaption{\quad\quad}
    \end{subfigure}
    \begin{subfigure}[t]{0.325\columnwidth}
    \centering
    \includegraphics[width=\columnwidth, trim={2.33cm, 0.5cm, 0.73cm, 0}, clip]{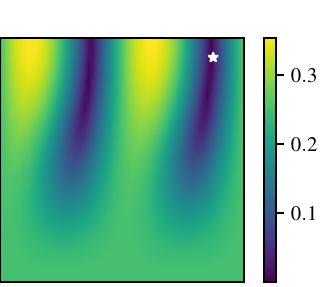}
    \subcaption{\quad\quad}
    \end{subfigure}
    \begin{subfigure}[t]{0.325\columnwidth}
    \centering
    \includegraphics[width=\columnwidth, trim={2.33cm, 0.5cm, 0.73cm, 0}, clip]{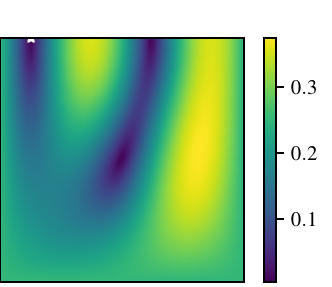}
    \subcaption{\quad\quad}
    \end{subfigure}
    \caption{Three unique probability landscapes of basis states for a $4$-qubit $3$-regular graph MaxCut instance. The repetition of these landscapes is related to the Hamming weight of the basis state bitstring. (a) Hamming weight $0$ or $4$; (b) Hamming weight $1$ or $3$; (c) Hamming weight $2$.}
    \label{fig:basis-state}
\end{figure}

A quantum state can be described by the amplitudes of the basis states: 
\begin{align}
    \ket{\Psi} = \sum_{x\in \{0,1\}^n} \alpha_x \ket{x},
\end{align}
where $n$ is the number of qubits and $\alpha_x$ is the amplitude of basis state $\ket{x}$. Upon measuring, the state collapses to one of the basis states with probability $|\alpha_x|^2$. This probability can be estimated by preparing and measuring the state many times. The expectation value used in VQAs uses such an estimation. The expectation value of a Hamiltonian given a state can be expressed as a weighted sum of each basis state's associated cost values:
\begin{align}
    \braket{\Psi(\bm{\theta})|\bm{H}|\Psi(\bm{\theta})} &= \sum_{x\in \{0,1\}^n} |\alpha_x(\bm{\theta})|^2 \braket{x|\bm{H}_x|x}\\
    &= \sum_{x\in \{0,1\}^n} |\alpha_x(\bm{\theta})|^2 C_x,
\end{align}
where $C_x = \braket{x|\bm{H}_x|x}$ is the cost value for solutions $x$ encoded in the Hamiltonian. Besides looking at the landscape of the expectation value, we can investigate the probability landscape of specific states. In particular, for classical optimization problems, there are one or multiple basis states that correspond to the optimal solution(s). Oftentimes, the goal is to find the parameters that generate optimal basis states with a high probability. In this case, the expectation value can be misleading, as parameters that produce high-probability suboptimal solutions can yield an expectation value smaller than those that lead to the optimal solution mixed with bad solutions. Thus, it is important to look at the probability landscapes of the optimal and suboptimal solutions to gain insights into initialization and optimizer configurations for similar problem settings.

Another scenario where basis state landscapes are useful is to investigate symmetries encoded in the problem. For example, a 4-qubit MaxCut instance has $16$ basis states as solutions. We can immediately see that the symmetry of variable mapping marks half of the solutions as duplicates. Nonetheless, the number of unique probability landscapes is only three, as shown in~\cref{fig:basis-state}, which is related to the Hamming weight of the corresponding bitstring. Such reflections of the symmetry in the original problem on state landscapes can be analyzed for VQA development.

\section{Conclusion}
In this paper, we propose a tensor-completion-based approach for reconstructing local cost landscape of VQAs. Our approach takes advantage of the low-rank property of local VQA landscapes and represents them in the tensor network format, which achieves exponential space reduction compared to the dense tensor representation. Our approach avoids exponential growth of required samples when the resolution increases, enabling high-resolution reconstruction with low overhead. Additionally, we show two examples of how landscapes are useful in developing VQAs. 

We highlight a few directions for future work. From the reconstruction algorithm perspective, it is worth investigating the properties of specific VQAs to further reduce the sampling cost of tensor completion, such as better strategies of the initial sampling and more customized tensor network structures for functional approximation of quantum algorithms. From the application perspective, it would be interesting to apply the landscape reconstruction techniques to study the impact of quantum noise and to perform error mitigation.

\small{
\section*{Disclaimer}
This paper was prepared for informational purposes with contributions from the Global Technology Applied Research center of JPMorganChase. This paper is not a product of the Research Department of JPMorganChase or its affiliates. Neither JPMorganChase nor any of its affiliates makes any explicit or implied representation or warranty and none of them accept any liability in connection with this position paper, including, without limitation, with respect to the completeness, accuracy, or reliability of the information contained herein and the potential legal, compliance, tax, or accounting effects thereof. This document is not intended as investment research or investment advice, or as a recommendation, offer, or solicitation for the purchase or sale of any security, financial instrument, financial product or service, or to be used in any way for evaluating the merits of participating in any transaction.
}
% \par\vspace*{\fill}

\bibliographystyle{Bib/IEEEtran}
\bibliography{Bib/main}
\end{document}